# Electrostatic Catalysis of a Click Reaction in a Microfluidic Cell

Semih Sevim[1], Carlos Franco[1], Albert C. Aragonès[2], Nadim Darwish[3], Donghoon Kim[1], Rosaria Anna Picca[4], Bradley J. Nelson[1], Salvador Pané[*,1], Ismael Díez-Pérez[*,5], Josep Puigmartí-Luis[*,2,6]

[1]*Institute of Robotics and Intelligent Systems, ETH Zurich, Tannenstrasse 3, CH-8092 Zurich, Switzerland.*

[2]*Departament de Ciència dels Materials i Química Física, Institut de Química Teòrica i Computacional, University of Barcelona (UB), Diagonal 645, 08028 Barcelona, Spain*

[3]*School of Molecular and Life Sciences, Curtin University, Bentley 6102, Western Australia, Australia.*

[4]*Chemistry Department, University of Bari "Aldo Moro", via E. Orabona 4, 70125 Bari, Italy.*

[5]*Department of Chemistry, Faculty of Natural, Mathematical & Engineering Sciences, King's College London, Britannia House, 7 Trinity Street, London SE1 1DB (United Kingdom)*

[6]*Institució Catalana de Recerca i Estudis Avançats (ICREA), Pg. Lluís Companys 23, 08010 Barcelona, Spain*

[*]*Corresponding author. Email : vidalp@ethz.ch, ismael.diez_perez@kcl.ac.uk , josep.puigmarti@ub.edu*

**Abstract**

Electric fields have been highlighted as a *smart reagent* in nature's enzymatic machinery, as they can directly trigger or accelerate redox and/or non-redox chemical processes with stereo- and regio-specificity. In natural catalysis, controlled mass transport of chemical species in confined spaces is also key in facilitating the transport of reactants into the active reaction site. Despite the opportunities the above offers in developing strategies for a new, clean electrostatic catalysis exploiting oriented electric fields, research in this area has been mostly limited to theoretical and experimental studies at the level of single molecules or small molecular ensembles, where both the control over mass transport and scalability cannot be tested. Here, we quantify the electrostatic catalysis of a prototypical Huisgen cycloaddition in a large-area electrode surface and directly compare its performance to the traditional Cu(I)-catalyzed method of the same reaction. Mass diffusion control is achieved in a custom-built microfluidic cell, which enhances reagent transport towards the electrified reactive interface while avoiding both turbulent flow conditions and poor control of the convective mass transport. This unprecedented electrostatic continuous-flow microfluidic reactor is an example of an electric-field driven platform where clean large-scale electrostatic catalytic processes can be efficiently implemented and regulated.

**Main**

Advanced computational studies have long proposed electric fields as key driving forces of enzymatic catalysis[1]. Recognizing this importance, electric fields have been recently referred to as "smart reagents" in nature's chemical machinery[2]. In nature, exceedingly large electric fields have been experimentally found to be electrostatically generated in confined spaces, *e.g.*, in the enzymes' active sites[3], playing a crucial role in accelerating chemical reactions[4]. Theoretical efforts have extensively predicted reaction rate and selectivity enhancements via applying oriented external electric fields (OEEF) along specific directions in relation to the reaction coordinate[5]. The challenge of both detecting and generating oriented electric fields in a confined chemical reactor have hampered experimentation in this field. Nonetheless, some recent experimental efforts support the theoretical framework for electrostatic catalysis. Fried *et al.* demonstrated that the enzyme ketosteroid isomerase exhibits a large electrostatic field at its active site, whose magnitude is strongly linked to the enzyme's catalytic efficiency[6]. In work by Aragonès *et al.*, a nanoscale junction was exploited to orient an electric field along the reaction coordinate of a confined non-redox Diels-Alder reaction, which resulted in a 5-fold reaction acceleration under moderate applied electric fields[7]. Here, we translate the electrostatic catalysis concept into a large surface area with the aim of up-scaling the process for its use in practical applications. A prototypical non-redox Huisgen cycloaddition, conventionally catalyzed by copper(I) salts[8–14], is effectively catalyzed over a molecularly functionalized centimeter square electrode surface exploiting voltage-controlled electric fields at the electrode/solution interface as the sole catalyst. Following nature's inspiration, the above catalytic processes is conducted under controlled mass transport conditions via a confined microfluidic channel, where turbulent flow and radial forces are avoided. This electrostatic continuous-flow microfluidic reactor allows quantification of electric-field driven chemical catalysis on a large-area electrified interface. We believe this is a significant step towards the implementation of a clean electrostatic catalysis in chemical industry applications.

**Results and Discussion**

**Electrostatic Catalysis Experiments in a Microfluidic Cell.** To investigate the catalytic effect of an OEEF on a prototypical click reaction, we have chosen a standard azide-alkyne cycloaddition, which is typically attained with high yields using $Cu^+$-based catalyst[8–14], even on a functionalized solid support[15]. To this end, we functionalize a gold-coated glass surface with an azide-terminated molecule ((**1**) in **Fig. 1**, left/right panels) *via* thiol-gold (S-Au) covalent chemistry[16] (**Supplementary Information (SI) 2.1-2.2**), and introduce a solubilized ferrocene alkyne derivative ((**2**) in **Fig. 1**, left/right panels) via a flow of acetonitrile solution of **2**. Upon reaction is completed (**Fig. 1** central panel), the ferrocene exposed groups attached to the functionalized gold-coated electrodes allow for further assessment of the yield of the reaction *via* electrochemical cyclic voltammetry (CV)[17]. Note that the surface-confined azide groups force the reaction to occur within the electrical double layer (EDL) region that develops nearby the electrode surface in contact with the polar solvent (*vide infra*) under an applied voltage. To accomplish constant mass transport of the alkyne reactant (*i.e.*, avoid surface mass depletion), the functionalized electrode was integrated into a microfluidic cell where the microscale channel is defined by two gold-coated glass electrodes separated by *ca.* 250 μm, which defines a reaction area of 5.7 cm x 1.2 cm (**SI Fig. S1a-c** and **SI section 2.3**). Once assembled, this microfluidic cell allows for introducing a continuous flow of a solution of **2** between the two electrodes while avoiding both turbulent flow conditions and a poor control of the convective mass transport. A voltage applied between the two electrodes creates the OEEF over the reaction area (left panel in **Fig. 1**).

In a typical experiment, a 0.1 M solution of ethynylferrocene in acetonitrile is injected between the two functionalized gold electrodes of the microfluidic cell at a constant flow rate of 50 μl/min for a reaction period of 30 minutes, while a voltage difference between 0.5 – 2 V is applied between the two parallel electrodes (see **SI 2.4**). The voltage-induced generated OEEF is expected to align the polar acetonitrile solvent molecules close to both electrified electrode surfaces forming EDL capacitors[18,19]. The latter results in the localization of the applied OEEF near the electrodes' surface, where the azide-alkyne cycloaddition reaction takes place. When the 30-minute reaction time concludes, the microfluidic cell is flushed with pure acetonitrile at a constant flow rate of 200 μl/min for a period of 5 minutes to remove the unreacted ethynylferrocene. Successfully *clicked* ethynylferrocene reactant molecules to the azide-terminated electrode surface are then quantified *via ex situ* cyclic voltammetry of the redox ferrocene groups (see **SI 3.1**). The integrated area of the characteristic voltammetric peaks of ferrocene is proportional to the amount of surface-bound ferrocene units, which is used to determine the surface reaction yield[20]. We use this methodology to explore the catalytic process under both electrostatic and chemical reaction conditions (**Fig. 1**).

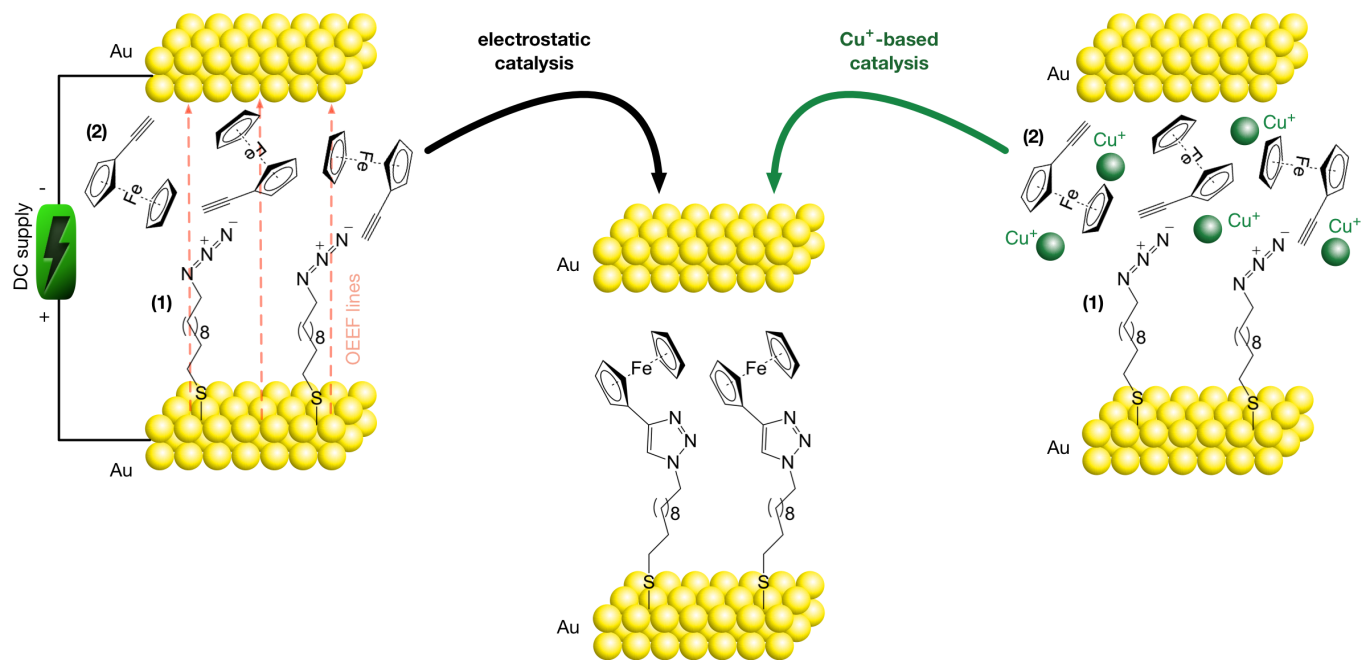

**Fig. 1. Catalysis of a click reaction in a microfluidic cell.** Schematic representations of oriented external electric-field (OEEF) in an electrostatic catalysis (left panel), and copper ($Cu^+$)-catalyzed click cycloaddition (right panel) in a confined microfluidic channel between two gold electrodes. The azide moiety is immobilized on the gold surface via thiol-gold chemistry, and the ferrocene alkyne derivative is flown continuously to avoid its depletion on the functionalized gold surface.

**OEEF Magnitude Effects on the Electrostatic Catalysis.** We first evaluate the efficiency of the electrostatically catalyzed azide-alkyne cycloaddition by comparing it to the same reaction under conventional $Cu^+$-based chemical catalysis conditions (**Fig. 2** and **SI Fig. S2**). A mixture of ethynylferrocene (0.1 M) and copper(I) iodide-triethyl phosphite (>20 mol% relative to the ethynylferrocene) in acetonitrile was injected between the two azide-terminated gold electrodes of the microfluidic channel using the same experimental conditions, namely, a constant flow rate of 50 µl/min for a period of 30 minutes, and in the absence of OEEF (*i.e.* no voltage applied), leaving $Cu^+$ as the only catalyzing agent for the click reaction (right panel in **Fig. 1**). The resulting cyclic voltammetric study demonstrates that the chemical $Cu^+$ catalysis results in a similar reaction yield as compared to the electrostatic catalysis performed at moderate electric fields generated under a 0.75 V of applied voltage (respectively, green and black curves in **Fig. 2a**). *Ex-situ* X-ray photoelectron spectroscopy (XPS) corroborates the existence of iron on the azide-terminated electrode surfaces, prepared under both electrostatic and $Cu^+$-based chemical catalysis (**SI Figs. S3-4**, respectively). The spectroscopic data also shows that the electrostatic catalysis yields much cleaner surfaces than $Cu^+$-based catalysis, the latter leaving residues from the used $Cu^+$ salt such as adsorbed iodide ions on the surface of the Au electrode (**SI Table 1**). Iodine contamination is also observed in consecutive CV cycles of samples prepared under $Cu^+$-based catalysis as an irreversible second oxidation peak corresponding to the $[I_3]^- \rightarrow I_2$ reaction[21] (**SI Fig. S5**). The absence of copper signal in the spectroscopic data suggests that this is eliminated from the substrate during the washing step.

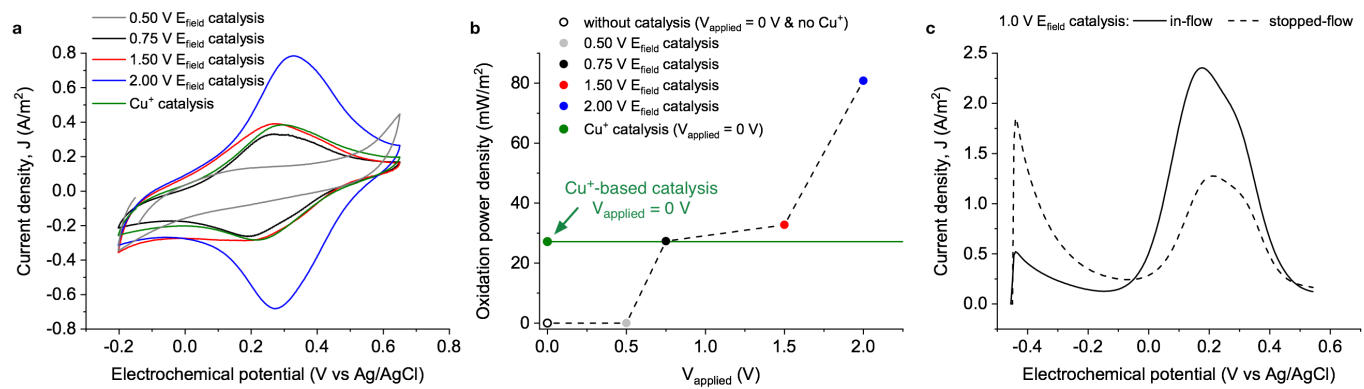

**Fig. 2. Performance of the electrostatic catalysis. a,** Cyclic voltammetry (CV) data showing effect of the applied voltage in the electrostatic catalysis, *e.g.*, 0.5 V (grey), 0.75 V (black), 1.5 V (red), 2 V (blue), and the comparison with $Cu^+$-based chemical catalysis (green). The two major peaks correspond to the oxidation-reduction of the ferrocene moieties attached to azide-terminated electrode via azide-alkyne cycloaddition. The CVs in **a** are obtained at a scan rate of 0.2 V/s. The CV analysis at different scan rates proves the redox couple comes from surface-bound species (see **SI Fig. S2**). **b,** Integrated areas under the anodic peaks of the CVs (positive peaks) in **a** for applied voltages 0.50 V (grey dot) 0.75 V (black dot), 1.5 V (red dot) and 2 V (blue dot), and for $Cu^+$-based catalysis at 0 V (green dot). Open circle is the control experiment performed under non-catalytic conditions (*i.e.*, $V_{applied}$ = 0 V and no presence of $Cu^+$). **c,** Square-wave voltammetry (SWV) of electrostatically catalyzed surfaces under continuous (solid line) and stop- (dashed line) flows in the microfluidic cell.

**Figs. 2a** and **b** show that the magnitude of the applied OEEF provides an exquisite control "knob" for the reaction yield. Different voltages ranging from 0.5 to 2 V were used to electrostatically catalyze the click reaction. CV results show net electrostatic catalysis for applied voltages ≥0.75 V (**Fig. 2a-b**). Electrostatic catalysis performed at lower voltages than 0.5 V resulted in CVs with no evident redox signal of ferrocene (grey curve in **Fig. 2a**) resembling those of bare gold and/or azide-terminated gold substrate (**SI Fig. S6**). Interestingly, at applied voltages of 1.5 V and ≥2 V, the reaction yield increased by 21% and 198%, respectively, over the values obtained for the $Cu^+$-based catalysis (**Fig. 2b**), indicating superior catalytic performance in the electrostatic version. It is important to note that all the experiments comparing reaction yields were performed using azide-terminated electrodes prepared in a single batch (*i.e.*, same metallization and functionalization batches). Due to batch-to-batch differences in the packing density of the surface functional groups, we observed slight variations regarding the shapes and areas of the voltammetric peaks as a result of disordered arrangements or supramolecular interactions between neighboring cyclopentadienyl groups[22,23]. Despite the latter, voltage-dependent catalytic effect was consistent across different batches (see results for a highly dense ferrocene layer under different OEEFs in **SI Fig. S7**).

**Mass-transport Effects on the Electrostatic Catalysis.** We conducted the electrostatic catalysis experiment under stagnant conditions, *i.e.,* stopped-flow, and compared it to the in-flow reaction shown in previous section. We used square-wave voltammetry (SWV) in this case to minimize the contribution of the capacitive current and increase the sensitivity (*i.e.,* peak integrability) of the low voltammetric signal for the stop-flow experiment[24]. The reaction yield for the in-flow conditions was 87% larger than the stop-flow under the same applied OEEF of 1 V (solid and dashed black curves in **Fig. 2c**). We observe that the enhanced mass-transport conditions under continuous flow improves final reactions yields across different applied OEEFs.

**OEEF Polarity Effects on the Electrostatic Catalysis.** The OEEF polarity is a crucial parameter in the mechanisms of electrostatic catalysis[18]. The OEEF polarity impact on the reaction is 2-fold; (1) it increases the probability of successful azide-alkyne cycloaddition by facilitating the proper alignment of the free ethynylferrocene reactant in solution with respect to the surface-bound azide group[25] (**Fig. 3a**), and (2), the particular OEEF polarity stabilizes the charge separated state of one of the resonance contributors (**Fig. 3b**) in the transition state (TS), thus reducing the reaction energy barrier[2,7,4,26–28,18,29] and resulting in higher reaction yield. Figure 3c shows the impact on the reaction yield upon OEEF inversion, *i.e.,* the area under the voltammetric peak is 73% smaller the negative OEEF polarity as compare to the positive (**Fig. 3c**), even at moderates applied OEEFs (0.75 V). These results evidence that while TS stabilization could be effective for the two opposed OEEF polarities, given the energy similarities among both most likely resonance contributors to the TS, the positive OEEF polarity (**Fig. 3a(*i*)**) enhances effective reactant collisions via electrostatic alignment of the free alkyne group in solution.

**Solvent Polarity Effect on the Electrostatic Catalysis.** We turned our attention to the polarity of the reaction medium as another key mechanistic ingredient of OEEF-based catalysis. Taking into account the EDL argument as the origin of the interfacial OEEF, a less polar solvent (*e.g.,* toluene) should hamper the OEEF localization near the electrode surfaces due to its much less efficient *shielding* of surface charges, leading to a thicker EDL spanning a larger distance from the electrode surface and resulting in lower OEEF magnitudes for the same applied voltages (**Fig. 3d(*i-ii*)**). In agreement with the above argument, we observe that the OEEF-based catalysis at 1.5 V voltage in toluene displays a 30% lower reaction yield than the conventional $Cu^+$ catalysis in the same solvent (solid red and green curves in **Fig. 3e**), whereas the OEEF-based catalysis at the same voltage in acetonitrile resulted in a comparable yield to the $Cu^+$ catalyzed reaction in toluene (see dashed red and solid green curves in **Fig. 3e**), on top of the higher observed yield when compared the $Cu^+$ catalyzed reaction in acetonitrile (**Fig. 2a-b**). These results demonstrate the EDL nature of the acting localized OEEF at the reactive electrode/solvent interface generating the electrostatic catalytic processes[30,31].

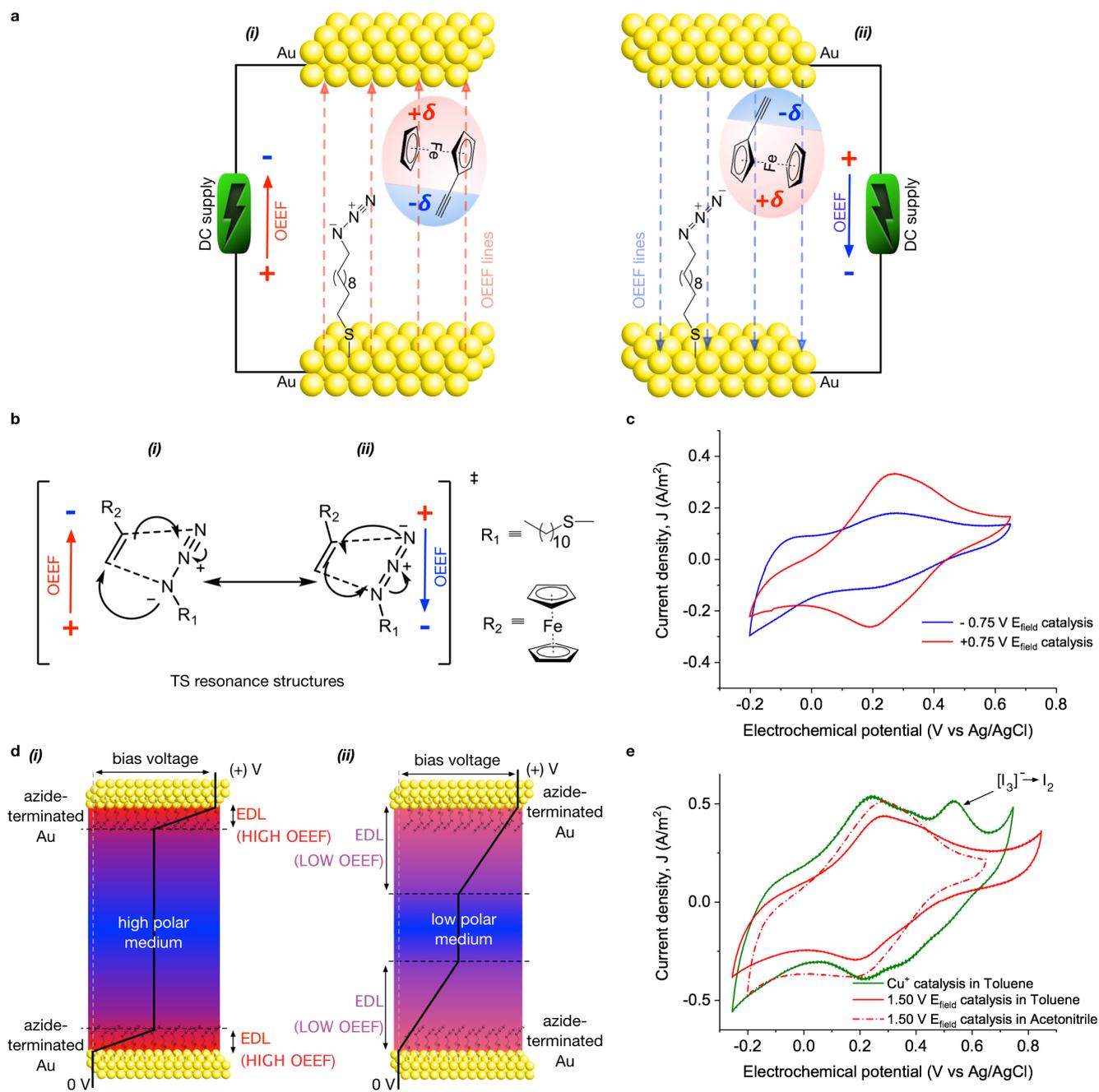

**Fig. 3. Effect of OEEF polarity in electrostatic catalysis. a,** Schematic representation of the OEEF lines and the alignment of the polar ethynylferrocene due to its dipole moment with *(i)* positive (+) and *(ii)* negative (-) applied voltage with respect to the azide-terminated working electrode. **b,** Most likely transition state (TS) resonance structures stabilized under opposite OEEF polarities (*e.g. (i)* positive (+) and *(ii)* negative (-) applied voltages. **c,** Cyclic voltammograms of chemical reactions induced using opposite voltage polarities of -0.75 V (blue) and +0.75 V (red). **d,** Schematic diagram showing EDL formation and corresponding voltage drop in *(i)* high polar (acetonitrile) and *(ii)* low polar solvents (toluene) respectively. **e,** Cyclic voltammograms of samples prepared in low polar medium (toluene) under Cu$^+$-based chemical catalysis (solid green line) and electrostatic catalysis at a voltage of 1.5 V (solid red line), as well as in a polar solvent (acetonitrile) under electrostatic catalysis applying the same voltage (dashed red line). Cyclic voltammograms presented in **c** and **e** were obtained at a scan rate of 0.2 V/s and 0.3 V/s, respectively.

**Quantification of the EDL-based OEEFs.** In view of the results from previous section (**Fig. 3e**), we attempted to quantify the magnitudes of the OEEFs generated within the EDL inducing electrostatic catalysis at the electrode/liquid interface. To this aim, we use voltage pulses followed by chronoamperometry (CA) measurements to determine the charge associated with the EDL formation[32] in different electrochemical arrangement as a comparative study (see **SI Fig. S8** and **SI section 3.2**). A step potential equivalent to the applied voltage ($V_{applied}$) used in the OEEF-based catalytic experiment is applied to the parallel electrodes in pure acetonitrile, and the resulting current transient is recorded. The measurements are repeated for consecutive charging and discharging cycles (**Fig. 4a**). Note that open circuit potential ($V_{OCP}$) were chosen to be the minimum (and initial) potential in the CA measurements in order to attain high reproducibility and less hysteresis[32,33] (see **SI section 3.2**). The resulting curves are then fitted to an exponential function using an *RC* circuit as a model to obtain the time constant of the EDL-based capacitor (*C*) (**Fig. 4b**) and, hence, the associated charge (*Q*) and the corresponding OEEF (*E*) across the EDL (**Table 1**).

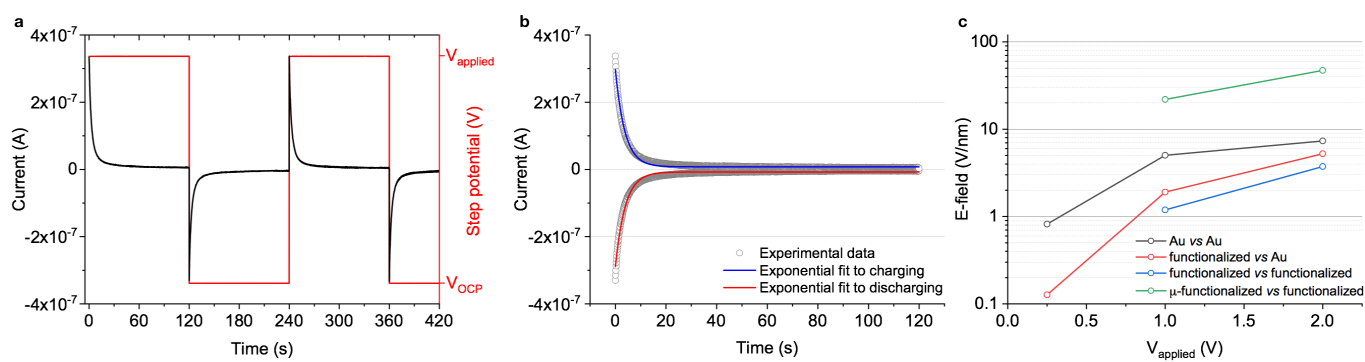

**Fig. 4. Characterization of the EDL-based OEEFs using an RC circuit model. a,** Representative chronoamperometry (CA) recordings showing charging and discharging currents (black) in consecutive applied step voltages (red) to the parallel electrodes immersed in pure acetonitrile. **b,** Exponential fits (blue and red lines) to the charging and discharging currents (grey circles) using an RC circuit as a model. **c,** Calculated electric fields from the RC fittings corresponding to different electrode configurations; (1) in a beaker containing pure acetonitrile and two parallel bare gold electrodes (Au *vs* Au) as a black, an azide-terminated gold electrode against a bare gold electrode (functionalized *vs* Au) in red, and two parallel azide-terminated electrodes (functionalized *vs* functionalized) in blue, and (2), in two parallel azide-terminated electrodes confined in the microfluidic cell (μ-functionalized *vs* functionalized) in green (see also **SI Fig. S8** for detailed schematic representations of electrochemical cell configurations used in CA measurements).

**Table 1. Calculation of the EDL-based OEEF magnitudes using an RC circuit model for constant voltages ranging between 0.25 V and 2 V in the different employed experimental configurations.**

| Configurations | | $V_{applied}$ (V) | $C$ (μF) | $Q$ (μC) | $E$ (V/nm) |
|---|---|---|---|---|---|
| Conventional electrochemical cell (electrode spacing = 7 mm) | 1st configuration: (Au *vs* Au) | 0.25 | 27.9 | 7.0 | 0.8 |
| | | 1.00 | 42.9 | 42.9 | 5.0 |
| | | 2.00 | 31.3 | 62.6 | 7.3 |
| | 2nd configuration: (functionalized *vs* Au) | 0.25 | 4.3 | 1.1 | 0.1 |
| | | 1.00 | 16.2 | 16.2 | 1.9 |
| | | 2.00 | 22.3 | 44.7 | 5.2 |
| | 3rd configuration: (functionalized *vs* functionalized) | 1.00 | 10.2 | 10.2 | 1.2 |
| | | 2.00 | 15.9 | 31.8 | 3.7 |
| Microfluidic cell (electrode spacing = 250 μm) | 4th configuration: (μ-functionalized *vs* functionalized) | 1.00 | 132.9 | 132.9 | 22.0 |
| | | 2.00 | 143.0 | 286.0 | 47.2 |

We investigated the magnitudes of the OEEFs for the different applied voltages ranging between 0.25 V and 2.0 V employed in the above electrostatic catalysis experiments (see black, red and blue lines in **Fig. 4c**). The CA experiments were done in both a conventional electrochemical and a microfluidic cell (see details in **SI section 3.2**). The former electrochemical setup provides fine control on the applied electrochemical potentials to both electrodes' interfaces and serves as the perfect benchmark for the quantification EDL charging/discharging process. The experiments performed in the conventional electrochemical cell using acetonitrile and bare Au electrodes results in large interfacial OEEFs of up to 7.3 V/nm at 2 V (**SI Fig. S8a*(i)*** and **Table 1**). The additional functionalization of either one or the two electrodes acted as an additional in series capacitance, which contributes towards charge shielding, hence decreasing the EDL-based OEEF generated under the same 2 V voltage down to 5.2 and 3.7 V/nm respectively (**SI Fig. S8a*(ii-iii)*** and **Table 1**). We then compare the above OEEFs to those equivalently generated in the microfluidic channel. To this aim, we filled our microfluidic cell with pure acetonitrile and applied voltage steps of 1 and 2 V to the two parallel electrodes, both decorated with the azide-terminated compound (**SI Fig. S8b**), The resultant OEEFs magnitudes appear to be significantly larger in the microfluidic channel (22 V/nm and 47 V/nm respectively for 1V and 2V, see **Table 1** and green line in **Fig. 4c**), owing to the confinement of the polar medium within the microfluidic channel, and facilitating successful electrostatic catalysis.

**Conclusion**

In summary, we have presented the first example of an electrostatic microfluidic-based approach for large scale electric-field driven catalysis under controlled mass transport. We demonstrated that the applied OEEF between the two electrodes confining the microfluidic channel can be used as a *'knob'* to regulate the yield of a prototypical Huisgen cycloaddition reaction generated at a large-area electrode/liquid interface. The correlation of the reaction yield with both the magnitude and polarity of the applied OEEF evidence the electrostatic nature of the catalysis. We also showed the key role of EDL in confining OEEF at the reactive electrode/liquid interface using solvents of different polarities, which results in different EDL thicknesses under the same applied voltages, hence regulating the final OEEF magnitude. The quantification of EDL-based OEEFs proves that the confinement of electrodes in a microfluidic reactor under exceptional mass transport conditions facilitates high interfacial OEEFs, thus achieving more efficient electric-field driven catalysis. We anticipate that our electrostatic continuous-flow microfluidic reactor will serve as a platform to open new routes to electric-field driven chemical synthesis towards a clean, more sustainable chemical synthesis.

**Acknowledgements**

This work is supported by the European Research Council Starting Grant *microCrysFact* (ERC-2015-STG No. 677020) and Consolidator Grant *Fields4CAT* (ERC-2019-CoG No. 772391), the Swiss National Science Foundation (project no. 200021_181988), and grant PID2020-116612RB-C33 funded by MCIN/ AEI /10.13039/501100011033.

**Author contributions**

S.P., I.D.P. and J.P.L. conceived the idea. S.S. designed the microfluidic cell, planned and performed all experiments, analyzed and interpreted the data. N.D. helped conceived the idea and with the data analysis. S.S. and C.F. performed the electrochemical characterizations. A.C.A. and D.K. helped with some of the experiments and the electrochemical characterizations. R.A.P. performed the XPS measurements. S.S., S.P., I.D.P. and J.P.L. wrote the manuscript. All authors contributed to writing the final version of manuscript.

**Competing interests**

The authors declare no competing interest

**Supplementary Information**

**Table of Contents**

# 1 Supplementary information for materials and methods

## 1.1 Materials

Bis(11-azidoundecyl) disulfide 99% (HPLC), Ethynylferrocene 97%, Copper(I) iodide-triethyl phosphite, Tetrabutylammonium perchlorate ≥99.0% (for electrochemical analysis), Acetonitrile (HPLC grade) ≥99.9%, Toluene ACS reagent ≥99.5%, Ethanol (absolute) ≥99.8%, Acetone ≥99.8% were purchased Sigma-Aldrich (Sigma-Aldrich Chemie GmbH, Switzerland). Hydrogen peroxide solution >30% (w/w) in $H_2O$ and Sulfuric acid >95% were bought from Acros Organics (Fisher Scientific AG, Switzerland). Isopropyl alcohol >99.7% was purchased from VWR International AG (Switzerland). All reagents were used without additional purification.

## 1.2 X-ray photoelectron spectroscopy (XPS)

X-ray photoelectron spectroscopic characterization was performed with a PHI Versaprobe II spectrometer (Chanhassen, MN, USA) equipped with monochromatic Al Kα source (1486.6 eV, 200 µm spot size). Charge neutralization was constantly applied. First, survey scans were acquired on at least three points for each sample at a 117.4 eV pass energy (1.0 eV step) to identify all the elements present on the surface. Then, high-resolution spectra were registered at 58.7 eV pass energy (0.125 eV step) for quantification. Binding energy (BE) scale was corrected on $Au4f_{7/2}$ component at 84.0 eV. Multipak software (v. 9.9.0.8, ULVAC-PHI, Inc., Japan) was used for element quantification. Chemical composition is expressed in terms of surface atomic percentage (El%).

## 1.3 Electrochemical characterizations

All electrochemical measurements, *e.g.* cyclic voltammetry (CV), square-wave voltammetry (SWV) and chronoamperometry (CA), were performed with Autolab/PGSTAT204 (Metrohm AG, Switzerland) using NOVA 2.0 software. More detailed information regarding electrochemical characterizations can be found in **Supplementary Information 3**.

## 2 Experimental methodology

### 2.1 Preparation of gold-coated glass slides

First, commercial microscope slides (75.5 mm x 25.5 mm, Thermo Scientific™ Microscope Slides, Ground 45°, Fisher Scientific AG, Switzerland) were drilled using a micro milling machine (MF 70, Proxxon GmbH, Germany) with a diamond tip to create holes (1 mm in diameter) for the inlet, outlet and electrical contacts. Specifically, the top glass slide has three holes, *i.e.* two holes for the liquid inlet and outlet to achieve continuous-flow of the reagent in the microfluidic cell, and one hole as an opening to take the electrical contact from the bottom electrode. On the other hand, the bottom glass slide has only one hole which is used as an opening to take the electrical contact from the top electrode (see **Supplementary Information 2.3**). Drilled glass slides were cleaned with acetone, isopropanol, and Milli-Q water, and dried with Nitrogen ($N_2$) gas prior to the metallization step. Then, they were coated with chromium (10 nm) as an adhesion layer and subsequently with gold (100 nm) as the electrode layer using an electron-beam evaporation system (Plassys Bestek, France).

### 2.2 Functionalization of gold-coated slides

Prior to the functionalization of the gold-coated glass slides with azide-terminated thiol, all of the glassware was cleaned using piranha solution (3:1 mixture of sulfuric acid, $H_2SO_4$ and hydrogen peroxide, $H_2O_2$) and vigorously rinsed with freshly prepared Milli-Q water to remove residual piranha solution. After the washing step with Milli-Q water, the gold-coated slides were sonicated in ethanol and dried with $N_2$ gas. The functionalization was performed by immersing the gold-coated glass slides in 1 mM ethanolic solution of Bis(11-azidoundecyl) disulfide for 24 hours at room temperature and under a $N_2$-saturated environment. Note that, to avoid the oxidation of the thiol, the ethanol was bubbled with $N_2$ gas (for ≥20 minutes) before the preparation of the solution used for the functionalization of the gold-coated glass slides. After the functionalization, the derivatized slides were immediately rinsed with ethanol to remove the residual of Bis(11-azidoundecyl) disulfide solution.

### 2.3 Design of the microfluidic cell

The microfluidic cell was designed as two separate parts (*i.e.* the top clamp and the bottom clamp, **Supplementary Fig. 1a**) using a 3D computer-aided design (CAD) software (SOLIDWORKS 2018) to mechanically clamp a spacer – *e.g. ca.* 250 µm-thick acrylonitrile butadiene rubber (NBR) sheet or *ca.* 100 µm-thick polytetrafluorethylene (PTFE) – between two gold-coated glass slides (*i.e.* electrodes), hence forming a reaction area confined between parallel electrodes **(Supplementary Fig. 1b)**. The top and bottom parts of the microfluidic cell were machined from aluminum. Specifically, the top part includes input/output ports for microfluidic connectors (10-32 Coned for 1/16" OD, IDEX Health & Science, LLC, USA) to connect PTFE tubing (1/16" OD, IDEX Health & Science, LLC, USA) and hence enable continuous-flow operations. Moreover, both the top and bottom parts have extra holes that match with the ones in the gold coated glass slides, and that are used to make the electrical contact to the electrode surfaces (**Supplementary Figs. 1a-d**). Once the microfluidic cell is assembled, it enables for not only the continuous-flow (with a syringe pump, neMESYS 290N, CETONI GmbH, Germany) of reagent solution between the functionalized working and counter gold electrodes but also the application of a bias voltage with a DC power supply (RND 320-KA3005D, RND Lab, Distrelec Group AG, Switzerland) to generate an oriented external electric field (OEEF) between these parallel gold electrodes.

## 2.4 Microfluidic experimental procedure using acetonitrile as a solvent (high polar medium case) [1]

*Electrostatic catalysis under continuous-flow conditions:* *(i)* The microfluidic cell was assembled by clamping a 250 µm-thick NBR spacer between the gold-coated glass slides (*i.e.* the azide-terminated working and counter electrode), which defines a confined reaction area (57 mm x 12 mm) between the two parallel electrodes (separation distance *ca.* 250 µm) (**Supplementary Fig. 1c**). *(ii)* The electrical contact for the top (or bottom) electrode was taken with a conductive wire going thrugh the holes on the bottom (or top) clamp, the bottom (or top) glass slide and the spacer until touching to the top (or bottom) electrode surface (**Supplementary Figs. 1a, c**). *(iii)* Prior to sending the reagent solution (*i.e.* 0.1 M ethynylferrocene in acetonitrile), the microfluidic cell was flushed with solvent (*i.e.* pure acetonitrile) with a flow rate of 200 µl/min for 5 minutes. *(iv)* Then, 0.1 M ethynylferrocene in acetonitrile was injected in the microfluidic cell (with a flow rate of 200 µl/min for ca. 2 minutes) to remove the solvent and quickly fill the reaction chamber with the reagent solution. *(v)* Once the reaction chamber was completely filled with the reagent solution, the flow rate was decreased to 50 µl/min and a constant DC bias voltage was applied to the parallel electrodes for a period of 30 minutes. Note that we also performed experiments with different constant DC bias ranging between 0 V – 2 V. *(vi)* After the 30 minutes reaction time under continuous-flow conditions, the DC bias was stopped, and pure acetonitrile was injected (200 µl/min for 5 minutes) in the microfluidic cell to remove the excess ethynylferrocene solution and clean the electrode surfaces to prevent any physiosorbed molecules or surplus reactants. *(vii)* Finally, the microfluidic cell was dissembled and the working electrodes were characterized in a separate electrochemical cell to *a posteriori* interrogate the yield of click reaction (**Supplementary Information 3.1**).

---

[1] Note that all reagent solutions were filtered prior to injection of microfluidic device to avoid any precipitation in solution.

***Electrostatic catalysis under stopped-flow conditions:*** The same experimental procedure as described above (see *Electrostatic catalysis under continuous-flow condition*) was followed except for step *(v)*. More specifically, while applying the bias voltage to the parallel electrodes the flow was stopped (*i.e.* flow rate was decreased to 0 instead of 50 µl/min).

***$Cu^+$-based chemical catalysis under continuous-flow conditions:*** The same experimental procedure as described above (see *Electrostatic catalysis under continuous-flow condition*) was repeated with a reagent solution containing $Cu^+$ as catalyzing agent – *i.e.* Copper(I) iodide-triethyl phosphite (>20 mol% relative to the ethynylferrocene) was added to 0.1M ethynylferrocene solution in acetonitrile – and no bias voltage was applied during the reaction step, (step *(v)*).

## 2.5  Microfluidic experimental procedure using toluene as a solvent (low polar medium case)[2]

***Electrostatic catalysis under continuous-flow conditions:***[3] *(i)* The microfluidic cell was assembled by clamping a 100 µm-thick PTFE spacer between gold-coated glass slides (*i.e.* the azide-terminated working and counter electrode) to define a confined reaction area (57 mm x 5 mm) between the parallel electrodes (separation distance *ca.* 100 µm) (**Supplementary Fig. 1d**). *(ii)* The electrical contact for the top (or bottom) electrode was taken with a conductive wire going thrugh the holes present on the bottom (or top) clamp, the bottom (or top) glass slide and the spacer until touching to the top (or bottom) electrode surface (**Supplementary Figs. 1a, d**). *(iii)* Prior to sending the reagent solution (*i.e.* 0.1 M ethynylferrocene in toluene), the microfluidic cell was flushed with solvent (*i.e.* pure toluene) with a flow rate of 200 µl/min for 5 minutes. *(iv)* Then, 0.1 M ethynylferrocene in toluene was injected in the microfluidic cell (with a flow rate of 200 µl/min for ca. 2 minutes) to remove the solvent and quickly fill the reaction chamber with the reagent solution. *(v)* Once the reaction chamber was completely filled with the reagent solution, the flow rate was decreased to 50 µl/min and a constant DC bias voltage (1.5 V) was applied to the parallel electrodes for a period of 30 minutes. *(vi)* After 30 minutes reaction time under the continuous-flow conditions, the DC bias applied was stopped, and a pure toluene solution was injected (200 µl/min for 5 minutes) in the microfluidic cell to remove the excess ethynylferrocene and clean the electrode surfaces to prevent the presence of any physiosorbed molecules or surplus reactants. *(vii)* Finally, the microfluidic cell was dissembled and the working electrodes were characterized in a separate electrochemical cell to *a posteriori* interrogate the yield of the click reaction (**Supplementary Information 3.1**).

---

[2] Note that all reagent solutions were filtered prior to injection of microfluidic device to avoid any precipitation in solution.

[3] Note that using toluene as a solvent requires to change the NBR spacer with a PTFE spacer due to solvent compatibility issues. Since the PTFE spacer (100 µm thick) is thinner and stiffer than NBR spacer (250 µm-thick), the width of reaction chamber was decreased from 12 mm to 5 mm in order to increase the sealing performance of spacer by increasing its surface area in contact with gold-coated glass slides.

***$Cu^+$-based chemical catalysis under continuous-flow conditions:*[4]** The same experimental procedure as described above (see *Electrostatic catalysis under continuous-flow condition*) was repeated. In this case, the reactant solution that was injected in the microfluidic cell contained Copper(I) iodide-triethyl phosphite (>20 mol% relative to the ethynylferrocene) and ethynylferrocene (0.1M) in toluene. The reaction was performed without applying a bias voltage (step *(v)*).

---

[4] Note that using toluene as a solvent requires to change the NBR spacer with a PTFE spacer due to solvent compatibility issues. Since the PTFE spacer (100 µm thick) is thinner and stiffer than NBR spacer (250 µm-thick), the width of reaction chamber was decreased from 12 mm to 5 mm in order to increase the sealing performance of spacer by increasing its surface area in contact with gold-coated glass slides.

# 3 Electrochemical characterizations

## 3.1 Cyclic voltammetry (CV) and square-wave voltammetry (SWV)

After the preparation of the samples (with either electrostatic or $Cu^+$-based chemical catalysis) in the microfluidic cell, the working electrodes (azide-terminated gold coated slides) were characterized in a separate electrochemical cell to *a posteriori* interrogate the yield of the click reaction by investigating the ferrocene attachment via azide-alkyne cycloaddition. The electrochemical cell was built in a glass beaker (50 ml) including the sample slide from the microfluidic experiment, a platinum (Pt)-coated glass slide and a silver wire as the working, counter and reference electrodes, respectively. The electrolyte solution (0.1 M tetrabutylammonium perchlorate in acetonitrile) was filled in electrochemical cell until the half of the area of the working electrode was immersed in the electrolyte solution, and then the electrochemical cell was bubbled with $N_2$ for 5 minutes. All the electrochemical measurements were performed with an Autolab/PGSTAT204 (Metrohm AG, Switzerland) using the NOVA 2.0 software. Prior to the electrochemical characterization of each sample, the electrolyte solution was changed with a fresh one, the reference and counter electrodes were cleaned with acetone and dried with $N_2$. Note that, the cyclic and square-wave voltammograms were recorded using a silver (Ag) wire as a reference electrode and then the electrochemical potential was corrected with respect to the Ag/AgCl standard reference electrode by measuring the open circuit potential of the Ag wire with respect to the Ag/AgCl standard reference electrode.

## 3.2 Chronoamperometry (CA) for characterization of localized electric field near electrode surfaces

In order to characterize the electric field near the electrode surface CA measurements were performed. To mimic the experimental conditions for the electrostatic catalysis, a two-electrode configuration was used in CA measurements. More specifically, the electrochemical cell was configured with only working and counter electrodes (*i.e.* without any reference electrode), and the pure acetonitrile solution is used as the electrolyte medium. Three different electrochemical cell arrangements were tested in a glass beaker (50 ml) (**Supplementary Fig. 8a**), *i.e. (i)* bare gold surface *vs* bare gold surface, *(ii)* azide-terminated gold surface *vs* bare gold surface and *(iii)* azide-terminated gold surface *vs* azide-terminated gold surface (as working *vs* counter electrodes), respectively. Moreover, the 4$^{th}$ electrochemical cell configuration was used to investigate the OEEF localized near electrode surfaces during the electrostatic catalysis experiments in our microfluidic cell (**Supplementary Fig. 8b**). To this end, the microfluidic cell was assembled by clamping a 250 µm-thick NBR spacer between two azide-terminated electrodes and filled with pure acetonitrile. After the microfluidic cell was completely filled with pure acetonitrile, the flow was stopped. All of the CA measurements were performed by starting from measured open circuit potential ($V_{OCP}$) and followed by applying a step voltage ($V_{applied}$, ranging between 0.25 V and 2 V) to the parallel electrodes for charging the formed electrochemical capacitor and subsequently discharging it at the open circuit potential ($V_{OCP}$). Note that $V_{OCP}$ were chosen to be the initial potential for CA measurements for high reproducibility and less hysteresis[1,2]. The charging/discharging response is associated with total capacitance on the electrode surface which is due to the double layer formation of pure acetonitrile and the azide-terminated functionalization layer on the gold surface (in case it is functionalized).

These above-mentioned configurations and experimental methodology allowed us to measure the current transient during the charging and discharging of the electrochemical circuit which can be modelled as a RC circuit to obtain the time constant[1]. More specifically, as charging progress current decreases exponentially (**Eq. *(1)***), while charge is increasing (**Eq. *(2)***) with following equations:

$$I(t) = \frac{V_{applied}}{R} e^{-t/RC} \qquad (1)$$

$$Q(t) = CV_{applied}(1 - e^{-t/RC}) \qquad (2)$$

where $I(t)$ and $Q(t)$ are respectively the current and charge at time $t$, $V_{applied}$ is the bias voltage applied to the parallel electrodes, $R$ is the resistance, $C$ is the total capacitance associated with charge accumulated on the electrode surfaces. During charging the initial current, $I(0)$ at $t=0$ is the maximum current, $I_{max}$ and could be obtained directly from the exponential fittings to CA measurements. Thus, the resistance of modelled circuit can be calculated with **Eq. *(3)***.

$$as \ t = 0, \quad I(0) = I_{max} = \frac{V_{applied}}{R} \Rightarrow R = \frac{V_{applied}}{I_{max}} \qquad (3)$$

The exponential fitting to the CA data during charging progress also gives us the time constant, $\tau$ of the modelled circuit which is used to calculate total capacitance associated with charge as in **Eq. *(4)***.

$$\tau = RC \Rightarrow C = \tau/R \qquad (4)$$

As time goes to infinity, the accumulated charge in the capacitor reaches its maximum value, $Q_{max}$ and could be obtained from **Eq. *(5)***. Finally, knowing the total charge associated with the electrochemical capacitor on the electrode surface allows us for the calculation of the localized electric fields near the electrode surface, $E_{max}$ with the Gauss's law (**Eq. *(6)***), where the $\varepsilon_0$ is the vacuum permittivity (8.85 x 10$^{-12}$ F/m).

$$\text{as } t \to \infty, \quad \lim_{t \to \infty} Q(t) = Q_{max} = CV_{applied} \qquad (5)$$

$$\text{as } t \to \infty, \quad \lim_{t \to \infty} E(t) = E_{max} = \frac{Q_{max}}{\varepsilon_0 A} \qquad (6)$$

# 4 Supplementary data

## 4.1 Supplementary figures

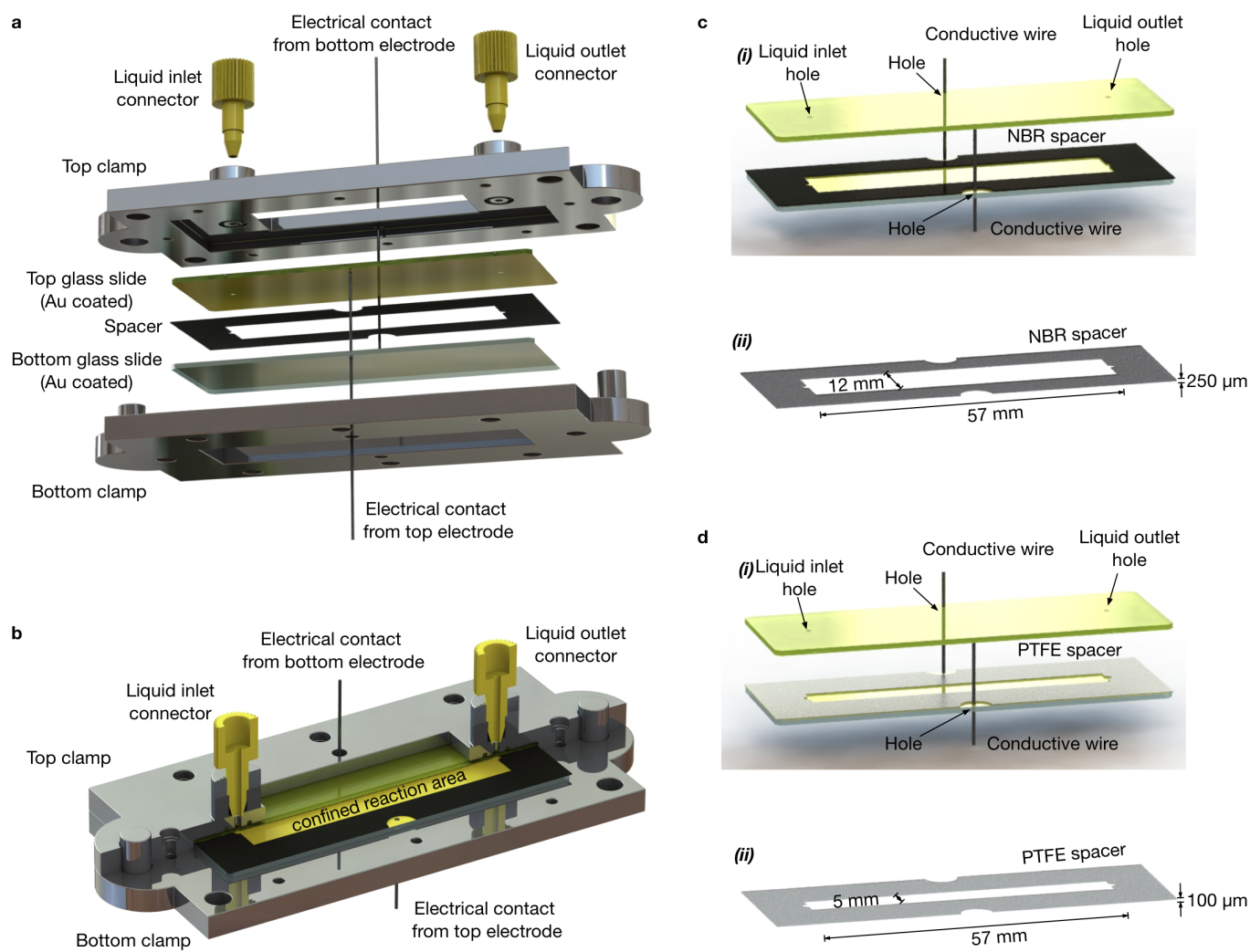

**Fig. S1. Design of microfluidic cell. a-b,** Schematic drawing showing **a,** split version of the microfluidic cell prior to its assembly to represent each part separately and **b,** assembled version of the microfluidic cell with a section-cut in the top part to demonstrate the confined reaction area formed with a spacer between two gold-coated electrodes. **c-d,** Isolated schematic view of *(i)* gold-coated glass slides showing the hole locations and *(ii)* spacer showing the confined reaction area dimensions used with **c,** acetonitrile and **d,** toluene, respectively. The top slide has three holes (*i.e.* two for the liquid inlet and outlet, and one for passing a conductive wire to make the electrical contact to the bottom slide), whereas the bottom slide has only one hole (for passing a conductive wire and make the electrical contact to the top slide).

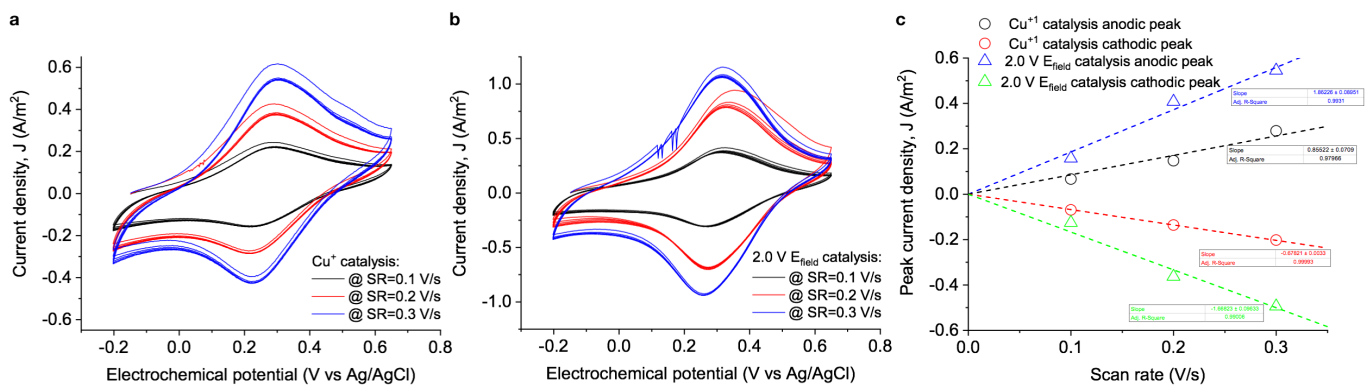

**Fig. S2. CV results with respect to different scan rates. a-b,** Multiple CV cycles at different scan rate (SR), such as 0.1 V/s (black), 0.2 V/s (red), 0.3 V/s (blue) performed on the sample prepared with **a,** $Cu^+$-based chemical catalysis and **b,** electrostatic catalysis with a bias voltage of 2 V. Stability in the oxidation-reduction peaks within multiple cycles indicates that ferrocene is chemically attached to the functionalized gold electrode via the azide-alkyne cycloaddition. **c,** Scan rate dependence of the peak current densities calculated from the cyclic voltammograms. Circles correspond to the anodic (black) and cathodic (red) peak current values calculated from CV data of samples prepared with $Cu^+$-based catalysis, whereas the triangles represent the anodic (blue) and cathodic (green) peak current values calculated from CV data of samples prepared with electrostatic catalysis under a bias voltage of 2 V. Color coded dashed lines represent the linear fitting to the corresponding data. The obtained linear relationship between the scan rate and the peak current densities proves that the redox couple comes from the surface-bound species[3].

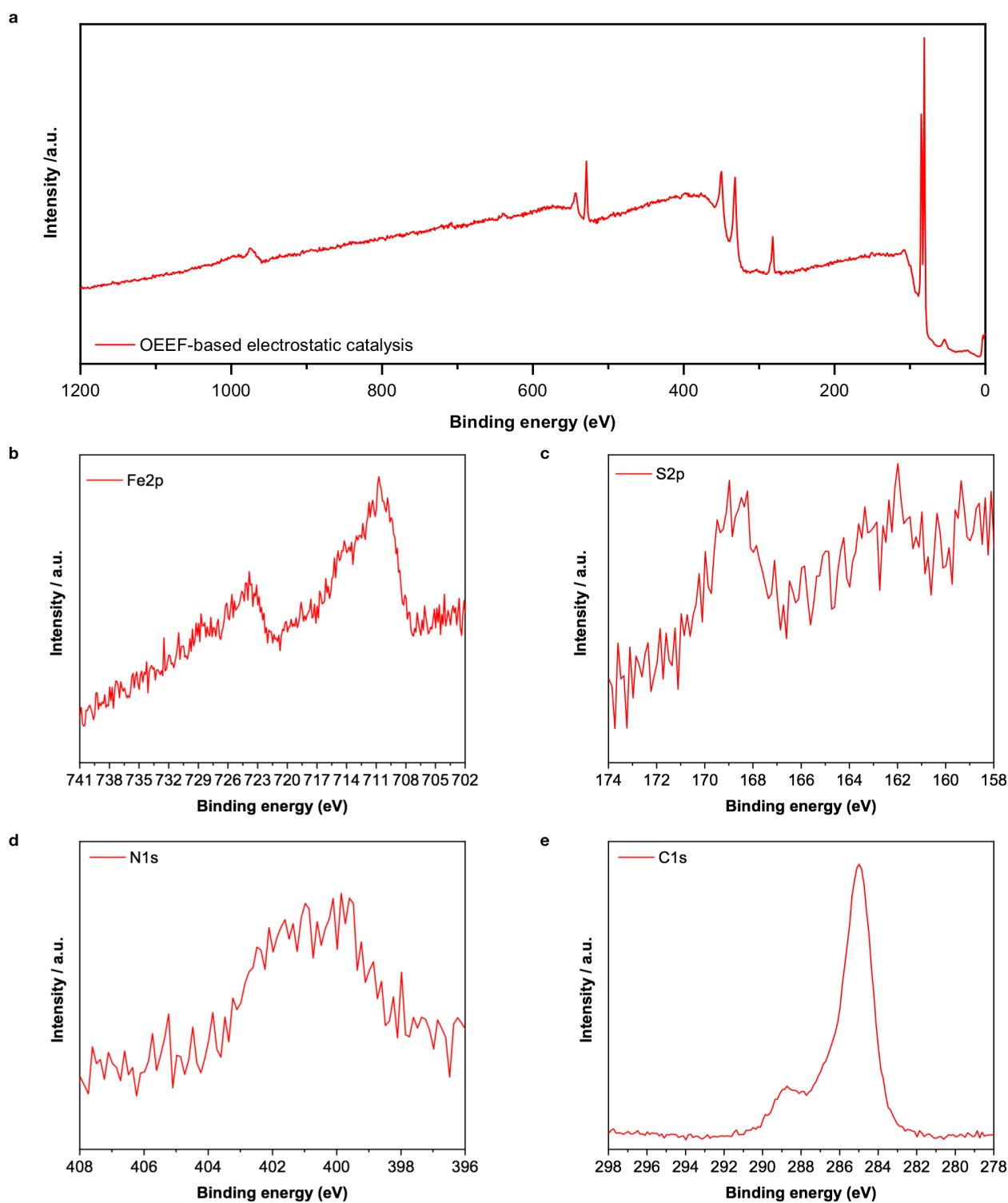

**Fig. S3. Chemical state analysis of the sample prepared with electrostatic catalysis. a,** Complete XPS spectrum of the sample prepared with electrostatic catalysis **b-e,** Corresponding high-resolution XPS spectrum showing the Fe2p, S2p, N1s, C1s regions respectively from **b** to **e**.

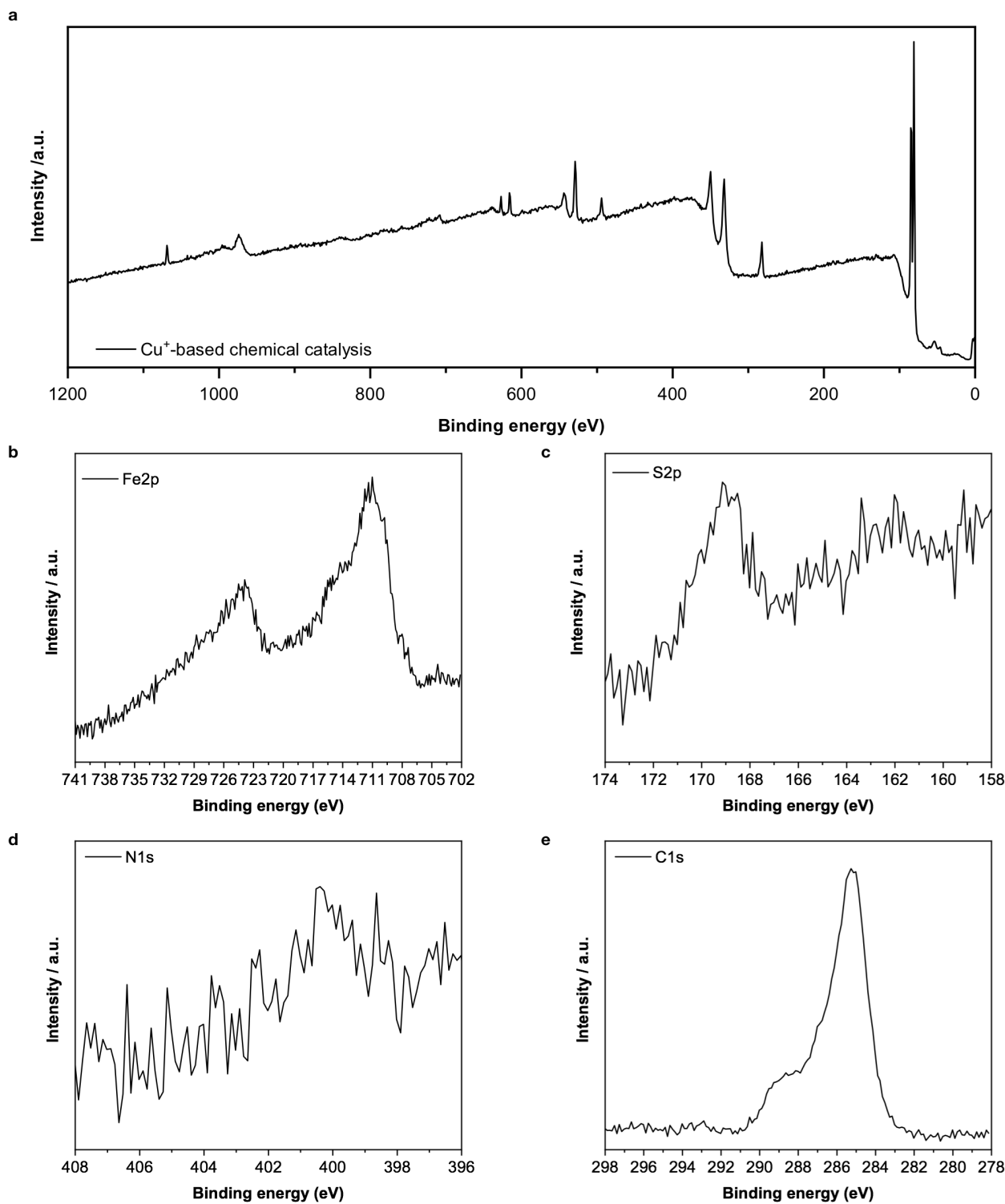

**Fig. S4. Chemical state analysis of the sample prepared with Cu$^+$-based chemical catalysis. a,** Complete XPS spectrum of the sample prepared with Cu$^+$-based chemical catalysis. **b-e,** Corresponding high-resolution XPS spectrum showing the Fe2p, S2p, N1s, C1s regions respectively from **b** to **e**.

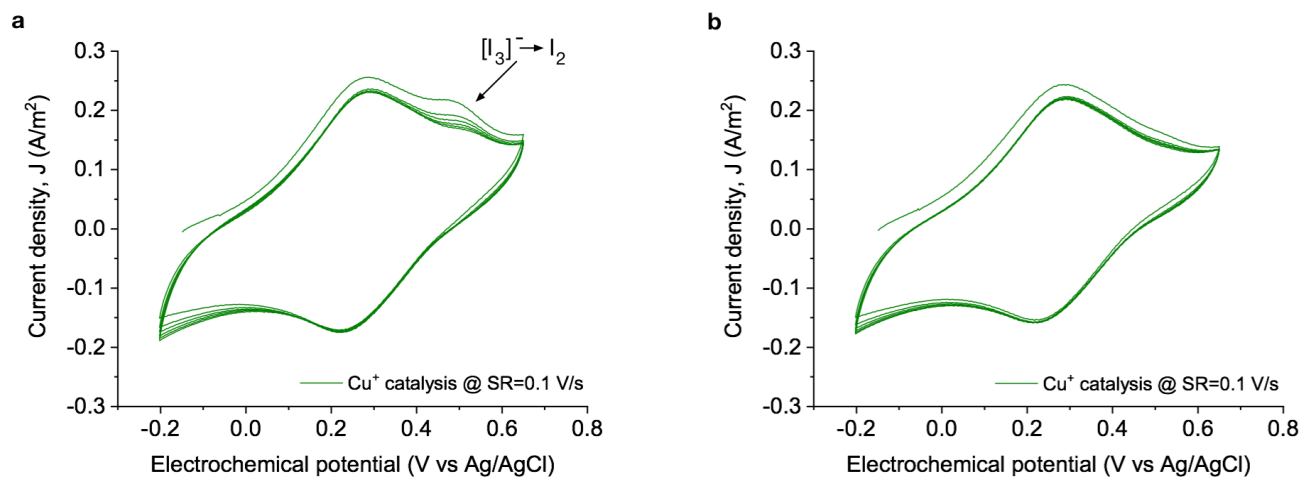

**Fig. S5. Surface contaminations during Cu$^+$-based chemical catalysis. a,** CV of the sample prepared with Cu$^+$-based chemical catalysis at a SR of 0.1 V/s showing an additional peak corresponding the oxidation of the precipitated iodide ([I$_3$]$^-$→I$_2$). Note that this oxidation peak is diminishing with consecutive cycles due to detachment of physiosorbed molecules from the electrode substrate, whereas the peaks that correspond to the oxidation and reduction of the chemically attached ferrocene are stable. **b,** CV of the same sample presented in **a**, after the physiosorbed materials are completely removed.

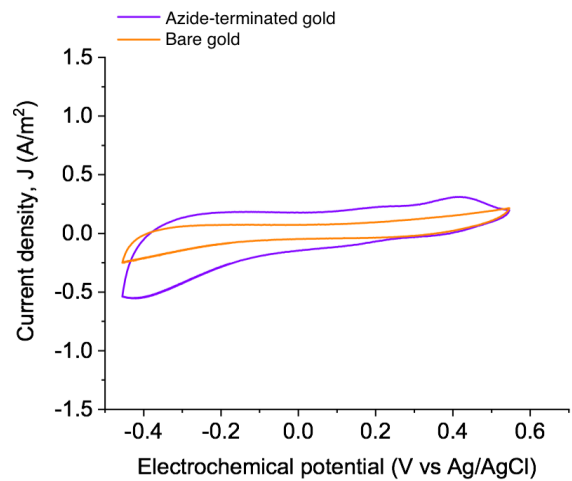

**Fig. S6. Control experiments for electrostatic catalysis.** CV of the bare-gold electrode (orange) and the functionalized (azide-terminated) electrode (purple) at a SR of 0.3 V/s.

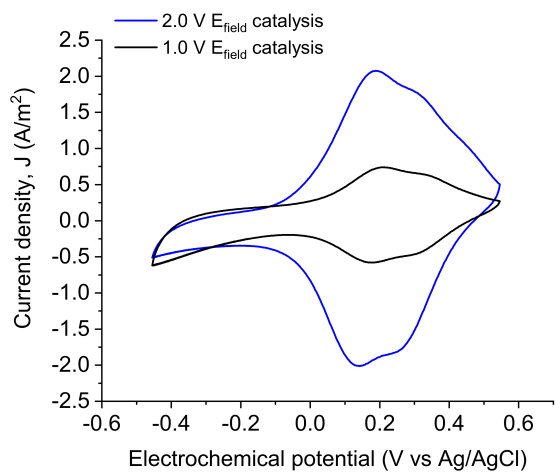

**Fig. S7. Effect of surface functionalization.** Oxidation-reduction of ferrocene attached to azide-terminated electrode via azide-alkyne cycloaddition catalyzed by OEEF at a bias voltage of 1 V (black) and 2 V (blue). Note that, these repetition experiments were performed using functionalized electrodes prepared in a new batch. Due to the likely differences in the formation of functional groups on the surface (*i.e.* disordered arrangement) and supramolecular interaction between cyclopentadienyl groups[4,5], the shape and area of the voltammetric peaks are slightly different than the ones presented in **Fig. 2a**. However, the effect of applied bias voltage is consistent with the results presented in the manuscript, *i.e.* 2V applied bias voltage results in a much higher yield of electrostatically catalyzed click reaction than the sample prepared with 1V bias voltage by employing functionalized electrodes from the same batch. The cyclic voltammograms are obtained at a scan rate of 0.3 V/s.

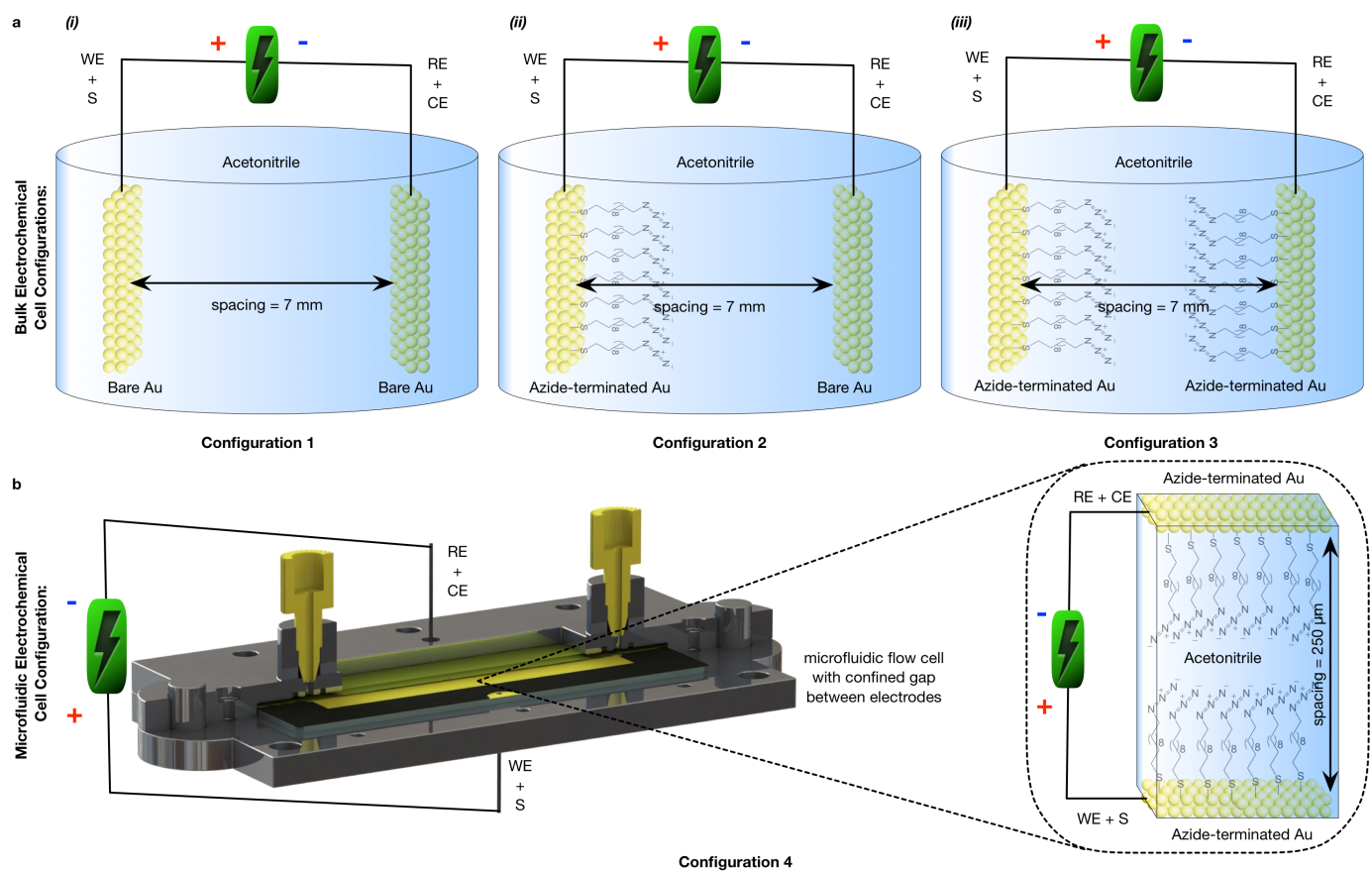

**Fig. S8. Electrochemical cell configurations for Chronoamperometry (CA) measurements. a,** Different electrode configurations in a beaker containing pure acetonitrile, such as *(i)* two parallel bare gold electrodes; *(ii)* an azide-terminated gold electrode against a bare gold electrode; *(iii)* two parallel azide-terminated electrodes. **b,** two parallel azide-terminated electrodes confined in microfluidic cell. All of the CA measurements were performed with two-electrode configuration (without using any reference electrode) and in pure acetonitrile (as an electrolyte medium). In the experiments where the glass beaker was used as an electrochemical cell, the distance between the working and counter electrodes was set to be 7 mm, whereas in the microfluidic electrochemical cell configuration the distance was set to 250 µm.

## 4.2 Supplementary tables

**Table S1. Surface chemical composition of the samples prepared via click reaction.** The atomic percentage of the elements detected by XPS of the samples prepared by either $Cu^+$-based chemical catalysis or electrostatic catalysis is reported, as mean value ± 1s (n = 3).

| Sample | C% | N% | O% | Na% | S% | K% | Fe% | I% | Au% |
|---|---|---|---|---|---|---|---|---|---|
| $Cu^+$ catalysis | 42.9±1.2 | 0.9±0.5 | 26.1±1.2 | 5.8±0.5 | 1.1±0.5 | ≤ 0.2 | 4.0±0.4 | 1.1±0.5 | 17.9±0.2 |
| Electrostatic catalysis | 51±5 | 3.2±0.4 | 24±3 | - | 1.0±0.5 | - | 2.4±0.4 | - | 18.4±1.5 |